\documentclass[reprint,
amsmath,
aip, apl 
]{revtex4-1}

\begin{document}


\title{Comment on the ``Decrease of the surface resistance in superconducting niobium resonator cavities by the microwave field''}

\author{A. Romanenko}
\email{aroman@fnal.gov}
\author{A. Grassellino}
\affiliation{Fermi National Accelerator Laboratory, Batavia, IL 60510, USA}%

\date{\today}
             
\maketitle

In a recent publication [Appl. Phys. Lett. 104, 092601 (2014)] Ciovati \textit{et al.} claim that: 1) thermal effects were disregarded in our original work~\cite{Romanenko_Rs_B_APL_2013}; 2) increase of $Q$ at $T=2$~K up to about $B\sim$100~mT in nitrogen doped cavities is just an extended low field $Q$ slope observed in non-doped cavities, which is furthermore attributed to the decrease of the ``BCS'' component of surface resistance. Here we show that both claims are wrong and the conclusions of Ciovati~\textit{et al.} are incorrect.

The first claim refers to our original paper~\cite{Romanenko_Rs_B_APL_2013} where the method of decomposition was first proposed and implemented. Contrary to Ciovati \textit{et al}, thermal effects were extensively analyzed in our paper based on the direct temperature mapping data and thermal simulations as can be seen in Fig.~2 therein. In particular the following conclusions have been clearly established: \textit{``We can conclude that (1) rf heating is negligible at lower temperatures and has no effect on the extracted $R_\mathrm{res}$; (2) the observed field dependence of $R_\mathrm{BCS}$ cannot be explained by the rf heating. It is important to reemphasize that rf heating does not introduce any new mechanisms of losses and has no effect on the residual resistance, the only thing it does is makes the temperature of the rf surface different from the bath temperature by an amount $\Delta T$, which depends on $B$ in a non-linear way. As we have shown the effect of such $\Delta T$ on our results is negligible.''} Thus the claim by Ciovati \textit{et al.} that ``thermal effects were disregarded'' is ungrounded and appears to be coming from the lack of full comprehension of our original work.

Since above $T_\mathrm{\lambda}=2.17$~K liquid helium becomes a normal fluid with the drastically lower thermal conductivity there is no surprise that if $Q(T)$ measurements to be extended above $T_\mathrm{\lambda}$ (like Ciovati \textit{et al.} have done) then the measured surface resistance exhibits a step increase due to the increase in the $\Delta T$ making the $R_\mathrm{BCS}$ contribution larger. However, contrary to what Ciovati \textit{et al.} state, the step-like increase of $R_\mathrm{s}$ \textit{above} $T_\mathrm{\lambda}$ does \textit{not} automatically imply that this contribution matters \textit{below} $T_\mathrm{\lambda}$. The temperature range, which was used for such deconvolution ($T<2.17$~K) in our original paper, was actually intentionally chosen to avoid any non-negligible thermal effects. 

Since Ciovati \textit{et al.} select the temperature range including temperatures above $T_\mathrm{\lambda}$ then thermal modeling and therefore relying on the values of Kapitza resistance and thermal conductivity taken from sample measurements elsewhere becomes necessary. This additional layer of indeterminacy (unnecessary below $T_\mathrm{\lambda}$) provides an extra source of systematic error, which negates the advantages of a larger range for pre-factor $A$ and gap $U$ fitting.

The second claim by Ciovati \textit{et al.} made from the beginning and throughout the article is that doping \textit{``significantly amplifies the low-field Q increase up to a factor of $\sim$2 and extends it to much higher fields $B=\mu_0 H \approx 60-90$~mT than what had been achieved previously''}. They further claim that it is the manifestation of the same effect and conclude: \textit{``Our analysis of experimental data suggests that the reduction of $R_\mathrm{s}$ at low frequencies $\hbar \omega \ll kT$ and temperatures $T \ll T_\mathrm{c}$ mostly comes from the current-induced broadening of the quasiparticle density of states.''}

However, abundant experimental evidence exists that the low field $Q$ slope is caused by the residual resistance, and not by the temperature-dependent ``BCS'' part. The obvious example is that the low field $Q$ slope remains when the temperature is lowered to $<$1.5~K where the thermally excited quasiparticle contribution (``BCS'') becomes negligible and the quality factor is no longer increasing upon further cooling - e.g. see Fig.~20 in Ref.~\onlinecite{Ciovati_JAP_2004} or Fig.~3 and Fig.~8 in Ref.~\onlinecite{Romanenko_JAP_2014}. The decrease of $R_\mathrm{s}$ with field at these temperatures is thus coming from the residual resistance. The same conclusion was also clearly supported by the decomposition method in the proper temperature range in our original paper~\cite{Romanenko_Rs_B_APL_2013}. On the contrary, the extended $Q$-rise in a wider field range observed in nitrogen doped cavities has a different origin and stems from the decrease of the ``BCS'' contribution as was also demonstrated in the past immediately after the discovery~\cite{Grassellino_SUST_2013}. 

We can further speculate that a possible source of the confusion between these two effects by Ciovati ~\textit{et al.} and attribution of them to the same origin is the systematic error in the fitting procedure introduced by the rf surface heating above $T_\mathrm{\lambda}$ described above.

In summary, the misconceptions and flaws in the experimental procedure we described above make the findings of Ciovati \textit{et al.} questionable and conclusions ungrounded.

%

\end{document}